# Photosensitive $SrMnO_3$

Arup Kumar Mandal[1], Aprajita Joshi[2], Surajit Saha[2], Binoy Krishna De[1], Sourav Chowdhury[1], VG Sathe[1], U. Deshpande[1], D. Shukla[1], Amandeep Kaur[3], D. M. Phase[1] and R.J. Choudhary[1]*

[1]*UGC-DAE Consortium for Scientific Research, Indore-452001, Madhya Pradesh, India*

[2]*Department of Physics, Indian Institute of Science Education and Research, Bhopal 462066, India*

[3]*Department of Physics, Indian Institute of Technology Bombay, Powai, Mumbai 400076, India*

**Abstract**

In recent years, photosensitive materials have been in huge demand because of their fascinating ability to convert absorbed photon energy to generate strain and henceforth tuning the physical properties. In this report we detect the photosensitive activity of $SrMnO_3$. Using the power dependent and temperature dependent Raman study with different laser sources having wavelengths across the optical band gap of $SrMnO_3$, we divulge the photosensitive character of $SrMnO_3$ thin films. Upon laser light illumination, Raman modes soften and softening further increases with increase in laser power. Similar kind of mode variation is observed with increasing temperature at fixed laser power. XAS in presence of laser illumination, reveals the change in crystal field splitting associated with mode softening.

*Corresponding author: ram@csr.res.in

Introducing strain in cross-functional materials and tuning their physical properties have been generating huge interests among the materials scientists in recent years [1-4]. Among several ways of inducing strain in multifunctional materials, photo induced strain has been attracting huge attention because of its easy switching ability between strain and relaxed state [5]. Among various perovskite based compounds, only few halides and oxides based systems have shown photosensitive distortion [6-21]. Particularly in transitional metal oxide series, this phenomenon is restricted in only few ferroelectric and multiferroic systems [7,13,20]. Microscopic origin of change in the physical properties of materials with light illumination is still not clear even for very common photostrictive materials [7]. This uncertainty arises because there is no one-to-one correspondence between the optical photon energy being used to induce photostrictive effects with the spin orbit interaction energy of 3d transition metal system in its ground state, although natural vibrational frequency of the perovskite systems is much smaller than the optical vibration frequency of light. In photosensitive system, absorbed photon energy causes changes in the materials ground state via secondary process such as photo-voltaic effect or inverse piezoelectric effect [7]. In such process, variation in local temperature upon photon absorption plays a crucial role, which can also lead to change in the materials ground state; for instance, local temperature assisted distortion can also modulate different physical properties of the system via variation in the electronic structure, available exchange interactions, hopping of the carriers etc.

In this study, we introduce $SrMnO_3$ as a photosensitive material. In transition metal perovskite series, $SrMnO_3$ is a very promising one. For the last decade, several interesting theoretical and experimental reports have been presented on thin film structures of $SrMnO_3$ [22-27], however, photo induced effects on structural and physical properties of this material are not yet explored. In this report we have performed power dependent Raman measurements (with 473 nm, 532 nm, and 633 nm laser source) across the optical band gap of 4H-SrMnO3 thin film grown on optically transparent quartz substrate. We have observed change in the *p-d* hybridization with the change in crystal field value when $SrMnO_3$ is illuminated with 532 nm and 633 nm light source.

## Experimental:

4H-$SrMnO_3$ (hexagonal strontium manganite) thin film was grown on optically transparent quartz substrate using pulsed laser deposition (PLD) technique (KrF excimer laser ($\lambda$ = 248 nm)) using dense pellet of single phase 4H-$SrMnO_3$ as target. The preparation procedure for 4H-$SrMnO_3$ is described elsewhere [25]. During thin film deposition, we maintained oxygen partial pressure at 120 mTorr and substrate temperature at 700° C. Structural single phase nature of the grown film was confirmed by gracing incident x-ray diffraction (GIXRD) using Bruker D8-Discover high resolution x-ray diffractometer with incident $\theta$ value 0.5° (see supplementary) [28], which also revealed (supplementary Fig. S1) its polycrystalline character. The diffraction pattern and calculated lattice parameters are shown in supplementary information (Fig. S1) [28]. Surface morphology and related information are extracted from AFM measurements, performed in contact mode (Fig. S1), as discussed in the supplementary information. [28]. X-ray photoemission spectroscopy (XPS) measurements were performed to confirm the chemical purity of the grown film using the Omicron energy analyzer (EA-125 Germany) with Al $K_\alpha$ (1486 eV) x-ray source (see supplementary information). Thickness of the grown film is calculated from X-ray reflectivity measurements and it is around 45 nm (Fig. S2) [28].

Power dependent Raman measurements were performed with 473 nm, 532 nm and 633 nm laser source at 300 K and 150 K. Power dependent and temperature dependent Raman measurements with 473 nm laser source was done by Horiba JY make HR-800 Raman instruments. Power dependent Raman spectra with 532 nm and 633 nm laser sources were collected with LabRAM HR Evolution Raman spectrometer in the backscattering geometry equipped with Peltier-cooled charge-coupled device (CCD) as detector. The sample temperature was varied using Linkam heating stage (Model HFS600E-PB4). Soft x-ray absorption near-edge structure (XANES) experiments were performed across Mn-*L* edge and O-K edge of SMO film along with $MnO_2$ and $Mn_2O_3$ reference samples with and without 532 nm and 633 nm light illumination at room temperature in total electron yield (TEY) mode at beamline BL-01, Indus-2, RRCAT, India.

**Results and Discussion:**

Fig. 1 (a) represents UV-Vis spectra (in transmission mode) of the grown film. Calculated band gap (direct band gap calculated from Touc plot) of the grown film is 2.5 eV (~ 495 nm) (see supplementary information). Fig. 1 (b) represents room temperature Raman spectrum of the grown SMO film recorded with 473 nm laser source. All vibration modes are assigned according to space group symmetry and previous study [25,29,30]. In 4H-SMO, one-unit cell consists of 4 $MnO_6$ octahedra, which form two $Mn_2O_9$ bioctahedra by three face sharing oxygen. Two bioctahedra in a unit cell are connected through Mn-O-Mn 180° bonding [31]. Closer inspection reveals that the $Mn_2O_9$ bioctahedra consist of two types of oxygen; face sharing oxygen (O1) and corner sharing oxygen (O2) (see schematic Fig.2). In 4H-SMO, only face sharing oxygen (O1) are Raman active [25,30]. According to group theory, eight vibrational modes are allowed in 4H-SMO structure ($\Gamma = 2A_{1g} + 2E_{1g} + 4E_{2g}$). Out of eight vibrational modes, only six single phonon modes are prominent for 4H-SMO (see Fig.2 schematic). Details of the observed modes for 4H-SMO are described elsewhere [25]. In Fig. 1 (b), the $A_{1g}^{(2)}$ (Ag) mode at around 640 $cm^{-1}$ represents O1 displacement in $Mn_2O_9$ bioctahedra. In both power dependent and temperature dependent Raman study (see supplementary information), we observed softening of Ag mode at around 640 $cm^{-1}$ as well as Eg mode at 435 $cm^{-1}$. However, shifting of the Eg mode towards lower frequency site is quite lower than the Ag mode shifting. Some of the lower frequency Eg modes (382 $cm^{-1}$, 220 $cm^{-1}$) remain invariant with respect to laser power or temperature. For this reason, we have focused our analysis based on Ag mode variation with respect to different laser power and temperature with 473 nm, 532 nm and 633 nm laser source.

Fig.3 a1 represents power dependent variation of Ag mode with 473 nm laser source at 150 K. Fig.3 a2 represents 3D intensity contour of Fig. 3 a1 data. From both the 2D and 3D plots, it is revealed that with increasing laser power, Ag mode shifts towards lower frequency by up to 5.7 $cm^{-1}$. Besides the Ag mode softening, it is also clear from the 3D plot that the broadening of the Ag mode increases with increasing laser power. Figs. 3 b1 and 3b2 represent the power dependent variation of Ag mode position and its intensity contour plot respectively with 473 nm laser source at 300 K. (Full Raman spectra are available in Fig. S3 (b), (supplementary part)) [28]. From both Fig. 3 b1 and b2, it is clear that with increasing laser power, Ag mode softening takes place along with broadening, similar to the observations at 150 K. Figs. 3 c1 and c2 represent the variation of Ag mode with temperature recorded with 473 nm laser source at 0.7 mW, which also reveal Ag mode softening with increasing temperature. (Full Raman spectra are available in Fig. S3 (c), (supplementary part)) [28]. From room temperature to 533 K, Ag mode shifts by 6.4 $cm^{-1}$ towards lower frequency at fixed laser power.

We also carried out similar exercise with laser source 532 nm as well as 633 nm at different temperature values and varying laser power and observed the variation in the Ag mode. Fig. 4 represents the power dependent (at 150 K and 300 K) and temperature dependent full stokes and anti-stokes Raman spectra with 532 nm laser source. Similar mode softening is observed at room temperature as well as 150 K and overall change in frequency (Ag mode) between minimum to maximum laser power with 532 nm laser source is 7.9 $cm^{-1}$ at 300 K and 7.3 $cm^{-1}$ at 150 K and that at fixed 1.69 mW laser power it is observed to be 7.1 $cm^{-1}$ shift as the temperature is increased from 300 K to 700 K. (details of Ag mode softening with varying laser power and with varying temperature is highlighted in the supplementary information Fig. , S4 with contour plot). We also recorded Raman spectra with 633 nm laser source. Fig. 5 represents power dependent (at 150 K and 300 K) and temperature dependent full stokes, anti-stokes Raman spectra with 633 nm laser source. With 633 nm laser source, Ag mode softens by 10.1 $cm^{-1}$ at 150 K and 10.3 $cm^{-1}$ at 300 K between minimum to maximum laser power, whereas at fixed 2.87 mW laser power, it softens by 10.3 $cm^{-1}$ with increase in temperature from 300 to 700 K. Details of Ag mode softening with varying laser power and with varying temperature is highlighted in supplementary information Fig. S5 with contour plot. For both the laser sources, softening of Ag mode takes place with increasing laser power along with increasing broadening of mode similar to the temperature dependent Ag mode variation at fixed laser power. It is noted here that the local structure modification (softening of Ag mode) with 9 mW laser illumination (with 532 nm and 633 nm) is similar to 1.2 % substrate induced distortion (Softening of Ag mode) in this structure (when compared with the Ag mode softening with different substrate induced strain in the SMO film) [25].

Overall, Raman measurements reveal that the most intense Ag mode of 4H-SMO shifts towards lower frequency side with increasing laser power of either of the 473, 532, 633 nm laser source. The Ag mode exhibits similar variation in temperature dependent Raman study at fixed laser power with 473, 532, 633 nm laser source. These observations are suggestive of a local rise in temperature of 4H-SMO in power dependent Raman shift. For better understanding of the power dependent Raman shift with the effect of local temperature, we analyzed both the Stokes and anti-Stokes Raman line with different power and temperature using 532 nm and 633 nm. One can calculate the local temperature by the intensity ratio of stokes and anti-stokes line with the following equation [7].

$$\frac{I_S}{I_{AS}} = \frac{\left(\vartheta_{excitation} - \vartheta_{phonon}\right)^3}{\left(\vartheta_{exciton} + \vartheta_{phonon}\right)^3} exp^{\left[h\vartheta_{phonon}/_{KT}\right]}$$

Where $I_s$ = intensity of stokes line, $I_{AS}$ = intensity of anti-stokes line, $\vartheta_{excitation}$ = frequency of the excitation laser, $\vartheta_{excitation} - \vartheta_{phonon}$ = frequency of Raman anti-stokes and $\vartheta_{excitation} + \vartheta_{phonon}$ = frequency of Raman stokes line. Accordingly, we calculate local temperature for all the spectra taken with different power and temperature with 532 nm and 633 nm laser source and plot them together to probe the variation of Ag mode as shown in Fig. 6(a) and (b). We have also taken into account the laser power induced heating effect in temperature dependent measurements for each temperature scan, so all the data points in Fig. 6 are in thermal equilibrium. From the fig. 6 (a), it is clear that when the spectrum is recorded at 150 K with 532 nm at highest available power (16.10 mW), the local temperature increases up to 740 K. Trend of power dependent Ag mode shift at 150 K and 300 K is similar to the temperature dependent shift in Ag mode at constant laser power. Fig. 6 (b) represents the variation of Ag mode with increasing laser power and temperature at 633 nm laser source. With 633 nm laser, the power dependent Ag mode softening takes place with increasing power, and this trend is similar to the temperature variation of Ag mode at fixed laser power. In this case also, we have considered the effect of laser heating for temperature dependent measurements. Measurement at the highest laser power (9.39 mW) at room temperature increased local temperature to 690 K. Even though a similar trend is observed in power and temperature dependent Ag mode softening with respect to local temperature, a considerable deviation is observed with 633 nm laser source. We have also calculated mode softening with respect to laser power and temperature with 473 nm laser source, the variation looks similar to the 532 and 633 nm laser sources, (see Fig.S3 supplementary information). However, the 473 nm laser source, we are unable to calculate the local temperature effect with increasing laser power (We are unable to collect anti-stokes line by Horiba JY make HR-800 Raman instruments because of absence of notch filter).

It is to be noted that the mode softening with increasing laser power is not the same for all the 473 nm, 532 nm and 633 nm laser sources. It is observed that with the increase in wavelength, softening of the Ag mode becomes more prominent for the same power difference, i.e. the Ag mode softening in 633 nm laser source is higher than 532 nm laser source for same laser power (see Fig. S4, S5 in supplementary information) [28]. So, SMO becomes more photoactive towards higher frequency region (UV-VIS-IR). In our SMO film, the laser induced local temperature rise is high enough (for this kind of granular surface morphology). Generally high number of grain boundaries reduces the effect of local temperature, because most of the absorbed photon energy is absorbed by the free electrons at the grain boundary [32-37], which contribute to the photocurrent as well**.**

Fig.7. (a) represents the Mn *L* edge XAS spectra of the grown SMO film. $L_{3,2}$ peak arises due to transition from spin-orbit split Mn $2P_{3/2,1/2}$ to unoccupied Mn 3d state. In this figure 7(a), we show XAS spectra of SMO with $MnO_2$ and $Mn_2O_3$ reference samples. SMO Mn *L* edge XAS spectrum appears similar to $MnO_2$, which confirms $Mn^{4+}$ charge state in the grown film of H-$SrMnO_3$, confirming the oxygen stoichiometric nature of the grown film. The origin of main peak, pre-peak and different features are discussed elsewhere [25]. To visualize the effect of laser induced local strain on electronic properties of SMO film, we performed XAS (O-K edge) measurements in presence of 532 nm and 633 nm laser sources as shown in Fig. 7(b). The O-K edge spectra arise due to the transition between O 1s-to- O 2p, which is hybridized with Mn 3d states. Thus O-*K* edge contains information about O 2p and Mn 3d hybridization along with unoccupied state of Mn 3d. First two features in O-*K* edge are attributed to the hybridization of Mn 3d with O 2p. Higher energy features are due to the hybridization of O-2p with Sr4d and Mn4sp respectively. First two peaks in O-*K* edge are very sensitive with respect to local distortion mediated change in p-d hybridization or change in stoichiometry [25].

We illuminate our sample with 532 nm and 633 nm laser light at incident angle of 45° to the normal of the sample surface. To compare the changes in the electronic structure upon the laser illumination, we align first feature of O-K edge of all the three normalized spectra at same energy position and examine second eg↓ feature position. It is observed that while illuminating SMO film with laser source, the eg↓ feature shifts towards higher energy side. We fitted first two features of O-*K* edge and calculated crystal field energy of SMO film under laser illumination. Crystal field of SMO film with normal condition is 2.2 eV, while with 532 nm and 633 nm illuminations, crystal field values are 2.5 eV and 2.7 eV, respectively. Local crystal field and effective density of state of 4H-SMO are highly sensitive to local distortion mediated change in *pd* hybridization [24,25].

The substrate induced local distortion can modify the crystal field in 4H-SMO compound [24,25]. In our present study we also notice light induced changes in local structure (from the blue shift of Ag mode upon light illumination) modifying the crystal field and effective p-d hybridization in this compound.

So, far it is evident that the SMO thin film generates huge local temperature rise upon laser light illumination. However, it needs to be probed if all the absorbed photon energy is spent in the local heating, or some parts of absorbed photon energy is participating in electron transport mechanism! To address this issue, we performed photocurrent measurements with different laser source with different power, along with temperature dependent resistivity measurements. Fig.8 represents room temperature I-V plot of SMO film with 473 nm, 532 nm and 633 nm light source, with power less than 5 mW. No such photocurrent is observed for any wavelength with this low power laser, even at higher voltages (see inset i2.). With 473 nm light source (see inset i3) up to 8 mW laser power also, we do not observe any enhancement in photocurrent. However, with 533 nm and 633 nm laser sources, we observed enhancement in photocurrent as estimated with the help of I-V measurements at 20 mW laser power (see inset i1). From Raman study it is observed that SMO local temperature increases when high intense laser source is illuminated. Thus, the local temperature mediated change in resistivity may be the cause for the enhancement in the photocurrent in SMO film. For that purpose, we performed high temperature (up to 515 K) resistivity measurement of SMO film, and extrapolated the data for high temperature. Fig. 8 (inset i4) represents resistance (R) versus temperature (T) of grown SMO film, which reveals decrease in resistance with increasing temperature. So, increase in photocurrent with high power laser source is mainly due to decrease in resistivity with increase in local temperature.

**Discussion:** In this present study, we observed that $SrMnO_3$ interacts with a broad optical frequency range. Local structure of $SrMnO_3$ is also modified while interacting with light, resulting in changes in its crystal field and *pd* hybridization. From Raman stokes and anti-stokes ratio we observed huge local temperature enhancement during laser illumination. 532 nm laser illumination at 9.18 mW laser power causes increase in local temperature up to 740 K, while 633 nm laser illumination at 9.38 mW laser power increases the local temperature around 690 K. So, at comparable laser power illumination, red light induced a relatively smaller local temperature rise with respect to green laser. We have analyzed temperature dependent Raman spectra at a fixed laser power and power dependence at fixed temperature, which shows similar variations in vibrational modes. With 532 nm laser source illumination, power dependent mode variation at different temperature and temperature dependent variation of Raman mode at fixed power follow exactly the same path. On the other hand, with 633 nm red laser source, though softening of the Ag mode is observed with the laser power as well as temperature variation, the trend of softening with the laser power and temperature does not follow each other and is slightly deviated. The mode softening with 633 nm laser is relatively large with respect to 532 nm laser source. In photo conductivity measurements, we observed highest conductivity with 633 nm laser source. So possibly $SrMnO_3$ converts some portion of photon energy directly in to mechanical energy of 633 nm red light. Generally, photostriction in other perovskite systems, for example single crystal of $BiFeO_3$, takes place through the ultra-violate (UV) and visible range, where the band gap lies in the visible light energy range [38]. However, in the present study we observed the photo-sensing phenomena even for laser energy values lower than the band gap of SMO.

**Conclusion:** We have successfully grown stoichiometric 4H-$SrMnO_3$ thin film on optically transparent quartz substrate. From power dependent Raman measurements, it is confirmed that the local structure of this $SrMnO_3$ thin film is modified with increasing laser power, and local structure modification with 9 mW laser illumination (with 532 nm and 633 nm) is similar to 1.2 % substrate induced distortion in this structure. From our Raman stokes and anti-stokes measurements, we observed high local temperature rise, for instance the local temperature at 300 K is enhanced to 700 K when illuminated with 532 nm lasers source at 9 mW. In correlated perovskite system, this high temperature rise is rare, particularly for granular morphic structure. Mode variation with respect to power and temperature is very similar for 532 nm laser source, while it is slightly diverted for 633 nm laser source. At the same time, we observed maximum photocurrent with 633 nm light source. By definition, photostriction must be a non-thermal phenomenon, on this basis $BiFeO_3$ is considered as a photostrictive material. However, a very recent report claims that during laser illumination, local temperature (from stokes anti-stokes ratio) in $BiFeO_3$ also increases hugely [7], and domain structure variation in epitaxial $BiFeO_3$ with laser light at fixed temperature and with temperature variation at fixed laser light source are similar. Even giant photostrictive $SrIrO_3$ thin film also shows temperature dependent structural distortion. So local temperature measurements in all such cases are debatable issue [39,40]. In our report we propose $SrMnO_3$ as a photosensitive material. Even in a granular relaxed structure, it generates huge local temperature throughout entire optical range, and local distortion increases towards infrared regime. Our finding will help us to explore this system as a solar power heater, local thermal insulating materials for thermal energy storage (TES), noncontact infrared thermometer, and as a backlit display material in future.

**Acknowledgements:** Authors thankful to Dr. V. R. Reddy for GIXRD and XRR measurements. Authors are thanking to Dr. Subhabrata Dhar for photo conductivity measurements, Dr. R Venkatesh and Mr. Mohan Gangrade for AFM measurements and Mr. Rakesh Shah and A Wadikar for XAS measurements. Author acknowledge to Parveen, Suman and Satish for their technical support. Dr. S Saha acknowledges the funding from SERB (CRG/2019/002668), India. Aprajita Joshi acknowledges CSIR (CSIR File No.: 09/1020(0179)/2019-EMR-I) for the research grant.

**Figure Caption:**

Fig. 1 (a) Represents UV-Vis spectra of 4H-$SrMnO_3$ thin film on quartz (in transmission mode). (b) Raman spectra of 4H-$SrMnO_3$ thin film with 473 nm laser source at room temperature.

Fig. 2 Schematic of allowed vibrational mode of 4H-$SrMnO_3$.

Fig. 3 (a1), (a2) are the power dependent variation of Ag mode of SMO film at 150 K with 473 nm laser source; (a2) is the intensity contour plot of (a1). (b1), (b2) Power dependent variation of Ag mode at 300 K with 473 nm laser source. (c1), (c2) Temperature dependent variation of Ag mode with 473 nm laser source at fixed 0.7 mW laser power (full spectra are available in supplementary material).

Fig. 4 (a) Power dependent stokes and anti-stokes Raman spectra of SMO film at 150 K with 532 nm laser source. (b) Power dependent stokes and anti-stokes Raman spectra of SMO film at 300 K with 532 nm laser source. (c) Temperature dependent stokes and anti-stokes Raman spectra with 1.69 mW laser source (532 nm).

Fig. 5 (a) Power dependent stokes and anti-stokes Raman spectra of SMO film at 150 K with 633 nm laser source. (b) Power dependent stokes and anti-stokes Raman spectra of SMO film at 300 K with 633 nm laser source. (c) Temperature dependent stokes and anti-stokes Raman spectra with 2.87 mW laser source (633 nm).

Fig. 6 (a) Power dependent variation of Ag mode at 150 K and 300 K including local temperature, calculated from stokes and anti-stokes ratio along with temperature variation of Ag mode at fixed laser power including laser induced temperature effect for 532 nm green laser source. (b) Power dependent variation of Ag mode at 150 K and 300 K including local temperature, calculated from stokes and anti-stokes ratio along with temperature variation of Ag mode at fixed laser power including laser induced temperature effect for 633 nm red laser source.

Fig. 7 (a) Represents Mn *L* edge XAS spectra of grown SMO film with $MnO_2$ and $Mn_2O_3$ reference sample. (b) O *K* edge XAS spectra of SMO film with normal condition and with illumination of 532 nm and 633 nm laser source; inset figure represents crystal field variation of SMO film with normal and 633 nm laser illumination (right inset represents the illumination schematic) (i) Is the schematic of XAS measurements.

Fig. 8 Room temperature I-V in presence of 473 nm, 532 nm and 633 nm laser source below 5 mW laser power. Inset (i1) Room temperature I-V with 532 nm and 633 nm laser source with 20 mW laser power. Inset (i2) Zoom view of low power I-V at high voltage regime. Inset (i3) Room temperature I-V with 478 nm laser illumination at different laser power. Inset (i4) represents variation in resistance of SMO film with temperature along with extrapolate data.

**Reference:**

[1]. R. Ramesh and Nicola A. Spaldin. nature materials | VOL 6 | January 2007.

[2]. Jing Ma, Jiamian Hu, Zheng Li, and Ce-Wen. Adv. Mater. 23, 1062–108. (2011).

[3]. N. Balke, S. Choudhury, S. Jesse, M. Huijben, Y. H. Chu, A. P. Baddorf, L. Q. Chen, R. Ramesh and S. V. Kalinin. NATURE NANOTECHNOLOGY | VOL 4 | DECEMBER 2009.

[4]. A. F. Th. Hoekstra, A. S. Roy, T. F. Rosenbaum, R. Griessen, R. J. Wijngaarden, and N. J. Koeman. Phy. Rev. Lett. 86. 5346. (2001).

[5]. V. Iurchuk, D. Schick, J. Bran, D. Colson, A. Forget, D. Halley, A. Koc, M. Reinhardt, C. Kwamen, N. A. Morley, M. Bargheer, M. Viret, R. Gumeniuk, G. Schmerber, B. Doudin, and B. Kundys, Phy. Rev. Lett. 117 107403 (2016).

[6]. B. Kundys, M. Viret, D. Colson and D. O. Kundys. Nat. Matt. 9, 803-805 (2010).

[7]. Yi-De Liou, Yo-You Chiu, Ryan Thomas Hart, Chang-Yang Kuo, Yen-Lin Huang, Yuan-Chih Wu, Rajesh V. Chopedakar, Heng-Jui Liu, Arata Tanaka, Chien-Te Chen, Chun-Fu Chang, Liu Hao Tjeng, Ye Cao, Valanoor Nagarajan, Ying-Hao Chu, Yi-Chun Chen and Jan-chi Yang. Nat. Matt. 18, 580-587 (2019).

[8]. Jan-Chi Yang, Yi-De Liou, Wen-Yen Tzeng, Heng-Jui Liu, Yao-Wen Chang, Ping-Hua Xiang, Zaoli Zhang, Chun-Gang Duan, Chih-Wei Luo, Yi-Chun Chen, and Ying-Hao Chu. *Nano Lett.* 18, 12, 7742–7748 (2018).

[9]. Dan Daranciang, Matthew J. Highland, Haidan Wen, Steve M.Young, Nathaniel C. Brandt, Harold Y. Hwang, Michael Vattilana,MatthieuNicoul, Florian Quirin,8 John Goodfellow, Tingting Qi, Ilya Grinberg, David M. Fritz, MarcoCammarata, Diling Zhu, Henrik T. Lemke, Donald A. Walko, Eric M. Dufresne, Yuelin Li, Jo¨rgenLarsson, David A. Reis, Klaus Sokolowski-Tinten, Keith A. Nelson, Andrew M. Rappe. Phy. Rev. Lett. 108, 087601 (2012).

[10]. Chen Chen and Zhiguo Yi. Adv. Funct. Mater. 31, 2010706, (2021).

[11]. Tzu-Chiao Wei, Hsin-Ping Wang, Heng-Jui Liu, Dung-Sheng Tsai, Jr-JianKe, Chung-Lun Wu, Yu-Peng Yin, Qian Zhan, Gong-Ru Lin, Ying-Hao Chu & Jr-Hau He. Nat. Comm. 15108. (2017).

[12]. Fernando Rubio-Marcos, David Paez-Margarit, Diego A. Ochoa, Adolfo Del Campo, José F. Fernandez, and José E. García. *ACS Appl. Mater. Interfaces*, 11, 15, 13921–13926 (2019).

[13]. Vivek Dwij, Binoy Krishna De, Sumesh Rana, Hemant Singh Kunwar, Satish Yadav, Shikha Rani Sahu, R. Venkatesh, N. P. Lalla, D. M. Phase, D. K. Shukla, and V. G. Sathe. Phy. Rev. B. 105, 134103 (2022).

[14]. Yang Zhou, Lu You, Shiwei Wang, Zhiliang Ku, Hongjin Fan, Daniel Schmidt, Andrivo Rusydi, Lei Chang, Le Wang, Peng Ren, Liufang Chen, Guoliang Yuan, Lang Chen & Junling Wang. Nat. Commun. 7, 8 (2016).

[15]. Xiu Li, Chen Chen, Faqiang Zhang, Xintang Huang, and Zhiguo Y. Appl. Phys. Lett. 116, 112901 (2020).

[16]. Yuelin Li, Carolina Adamo, Pice Chen, Paul G. Evans, Serge M. Nakhmanson, William Parker, Clare E. Rowland, Richard D. Schaller, Darrell G. Schlom, Donald A. Walko, Haidan Wen & Qingteng Zhang. Scientific Reports 5:16650 | DOI: 10.1038/srep16650 (2015).

[17]. V. Juvé, R. Gu, S. Gable, T. Maroutian, G. Vaudel, S. Matzen, N. Chigarev, S. Raetz, V. E. Gusev, M. Viret, A. Jarnac, C. Laulhé, A. A. Maznev, B. Dkhil, and P. Ruello. Phy. Rev. B. 102, 220303(R) (2020).

[18]. Haidan Wen, Michel Sassi, Zhenlin Luo, Carolina Adamo, Darrell G. Schlom, Kevin M. Rosso & Xiaoyi Zhang. Scientific Reports 5:1509, (2015).

[19]. Subhajit Pal, Atal Bihari Swain, Pranab Parimal Biswas, and Pattukkannu Murugavel. Phy. Rev. Matt. 4, 064415 (2020).

[20]. Bingbing Zhang, Xu He, Jiali Zhao, Can Yu, Haidan Wen, Sheng Meng, EricBousquet, Yuelin Li, Chen Ge, Kuijuan Jin, Ye Tao, and Haizhong Guo. Phy. Rev. B. 100, 144201 (2019).

[21]. Xiaoliang Miao,aTing Qiu, Shufang Zhang, He Ma, Yanqiang Hu, Fan Baiaand Zhuangchun Wu. J. Mater. Chem. C, 5, 4931 (2017).

[22]. C. Becher, L. Maurel, U. Aschauer, M. Lilienblum, C. Magén, D. Meier, E. Langenberg, M. Trassin, J. Blasco, I. P. Krug, P. A. Algarabel, N. A. Spaldin1, J. A. Pardo, and M. Fiebig, Nat. Nanotechnol. 10, 661 (2015).

[23]. R. Guzman, L. Maurel, E. Langenberg, A. R. Lupini, P. A. Algarabel, J. A. Pardo, and C. Magen, Nano Lett. 16, 2221 (2016).

[24]. P. Agrawal, J. Guo, P. Yu, C. Hebert, D. Passerone, R. Erni, and M. D. Rossell. Phy. Rev. B. 94, 104101 (2016).

[25]. Arup Kumar Mandal, Anupam Jana, Binoy Krishna De, Nirmalendu Patra, Parasmani Rajput, V. Sathe, S. N. Jha, R. J. Choudhary, and D. M. Phase. Phy. Rev. B. 103, 195110 (2021).

[26]. Jun Hee Lee and Karin M. Rabe. Phy. Rev. Lett. 104, 207204 (2010).

[27]. Alexander Edström and Claude Ederer. Phy. Rev. Lett. 124, 167201 (2020).

[28]. Supplementary information

[29]. A. Sacchetti, M. Baldini, F. Crispoldi, P. Postorino, P. Dore, A. Nucara, C. Martin, and A. Maignan, Phys. Rev. B 72, 172407 (2005).

[30]. A. Sacchetti, M. Baldini, P. Postorino, C. Martin, and A. Maignan, J. Raman Spectrosc. 37, 591 (2006).

[31]. Rune Søndenå, Svein Stølen, and P. Ravindran. Phy. Rev. B. 75, 214307, (2007).

[32]. Jan Seide, Deyi Fu, Seung-Yeul Yang, Esther Alarco´n-Llado, Junqiao Wu, Ramamoorthy Ramesh, and Joel W. Ager III. PRL 107, 126805 (2011).

[33]. Sabine Körbel and Stefano Sanvito. Phy. Rev. B. 102, 081304(R) (2020).

[34]. Nastaran Faraji, Clemens Ulrich, Niklas Wolff, Lorenz Kienle, Rainer Adelung, Yogendra Kumar Mishra, and Jan Seidel. Adv. Electron. Mater. 2016, 2, 1600138

[35]. Subhajit Nandy, Pavana S. V. Mocherla, Kulwinder Kaur, Sanjeev Gautam, B. R. K. Nanda, and C. Sudakar. J. Appl. Phys. 126, 235101 (2019

[36]. Zhaodong Chu, Mengjin Yang, Philip Schulz, DiWu, Xin Ma, Edward Seifert, Liuyang Sun, Xiaoqin Li, Kai Zhu & Keji Lai. Nat. Commun.8, 2230 (2017).

[37]. Niladri Sarkar, Subhabrata Dhar and Subhasis Ghosh. J. Phys.: Condens. Matter 15, 7325–7335 (2003)

[38]. B. Kundys, M. Viret, C. Meny, V. Da Costa, D. Colson, and B. Doudin. Phy. Rev. B 85, 092301 (2012).

[39]. Anju Ahlawat and V. G. Sathe. J. Raman Spectrosc. 42, 1087-1094. (2011).

[40]. A. Laikhtman and A. Hoffman. J. Appl. Phys. 82 (1), (1997).

Fig.1

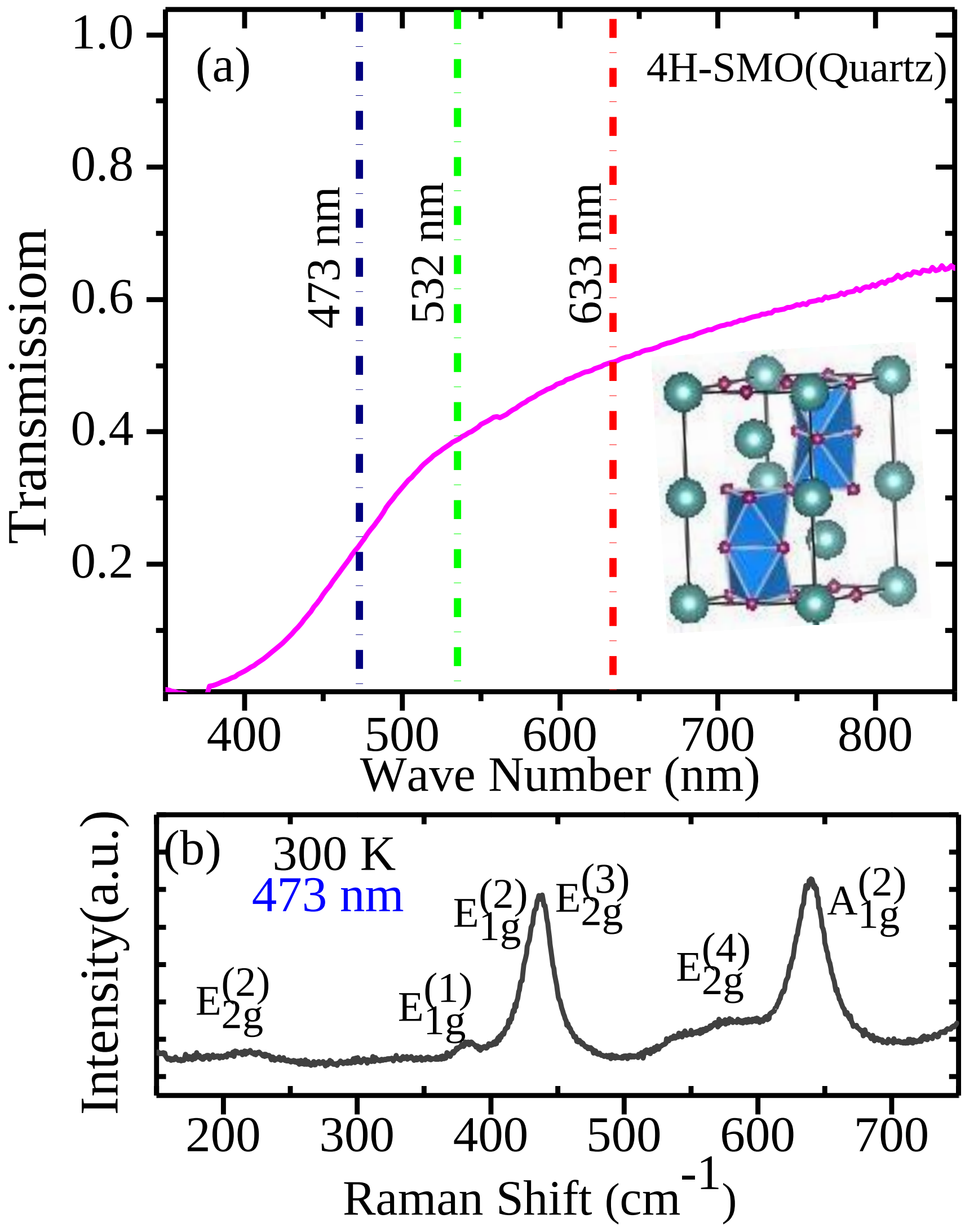

Fig.2

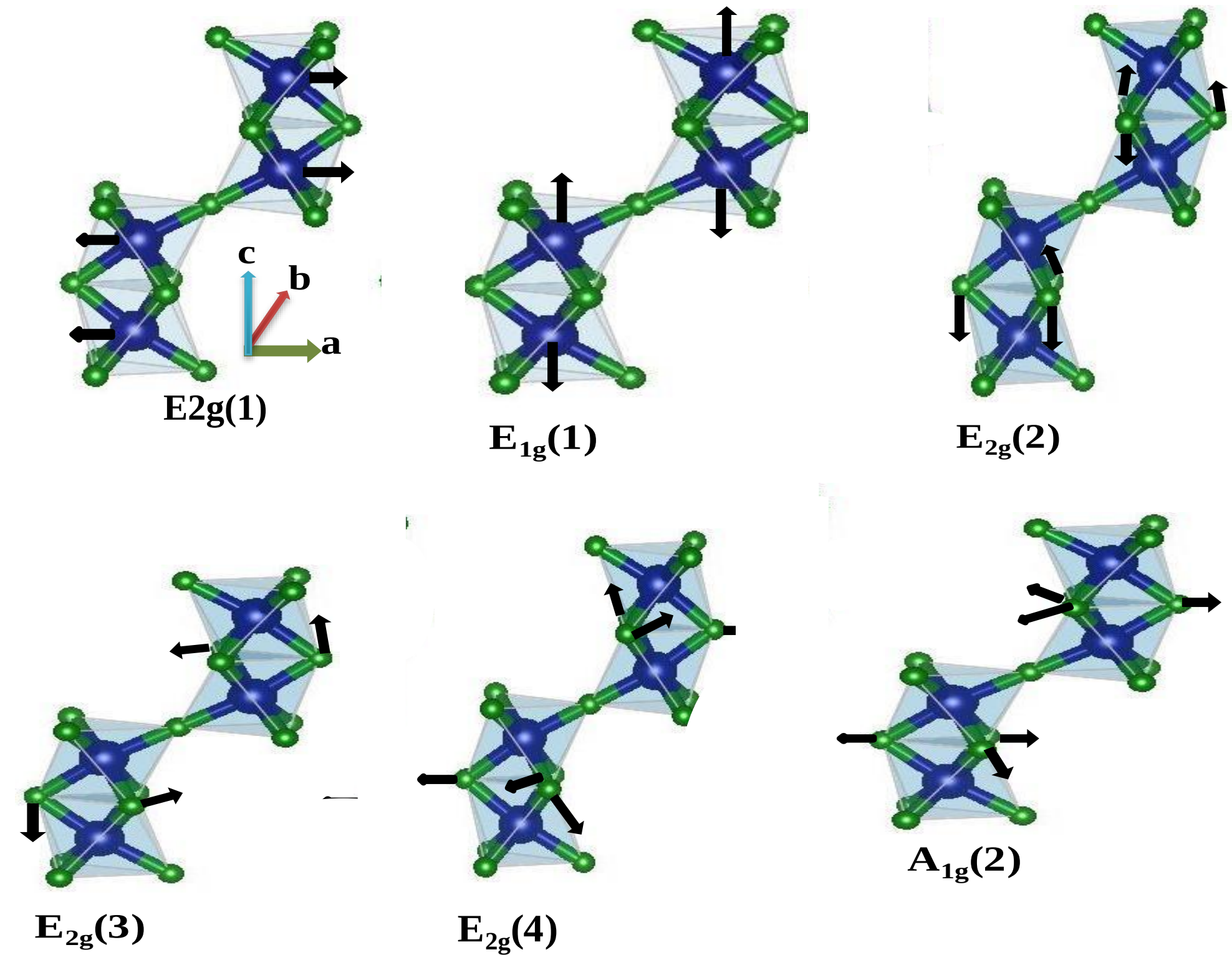

Fig.3

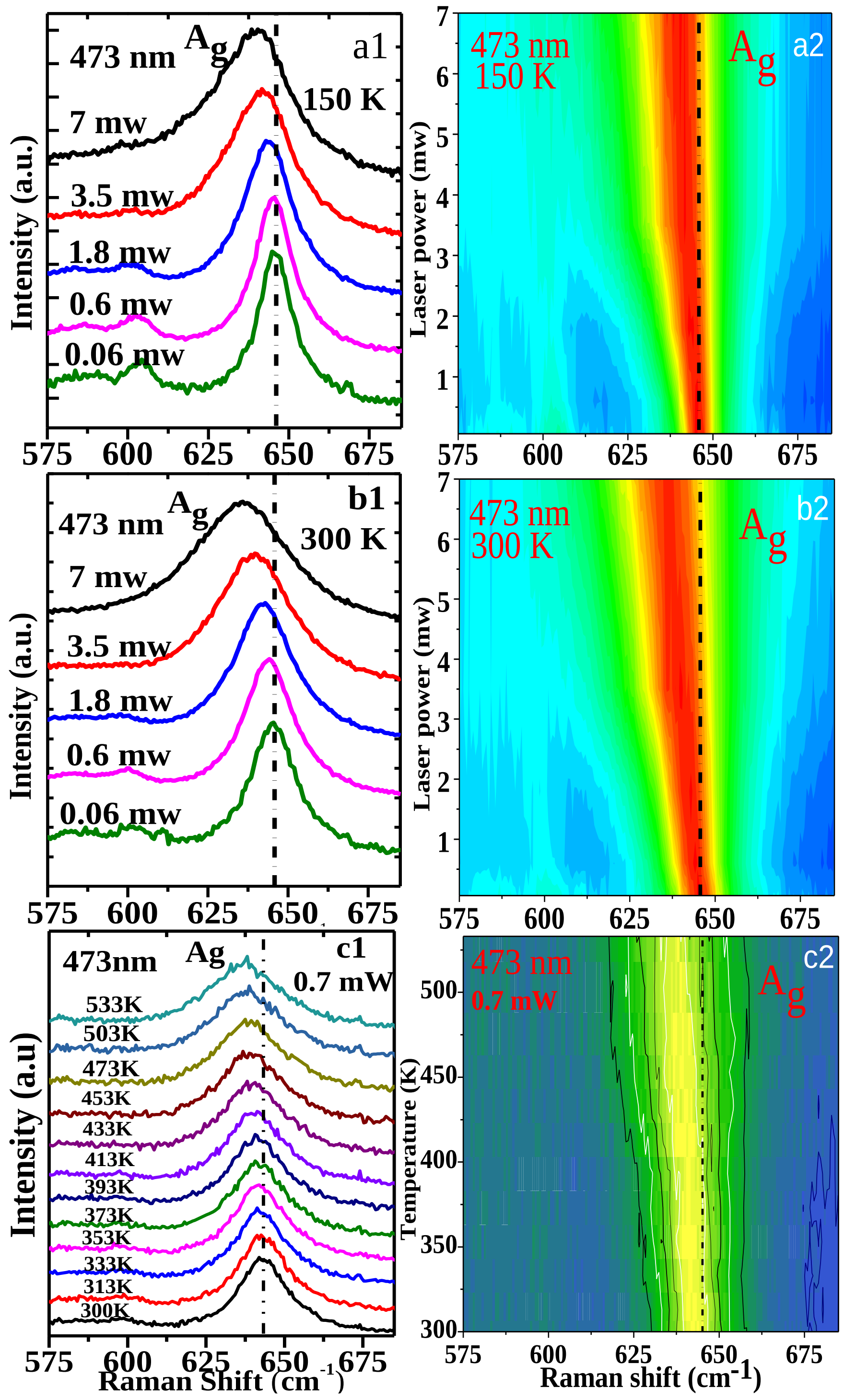

Fig.4

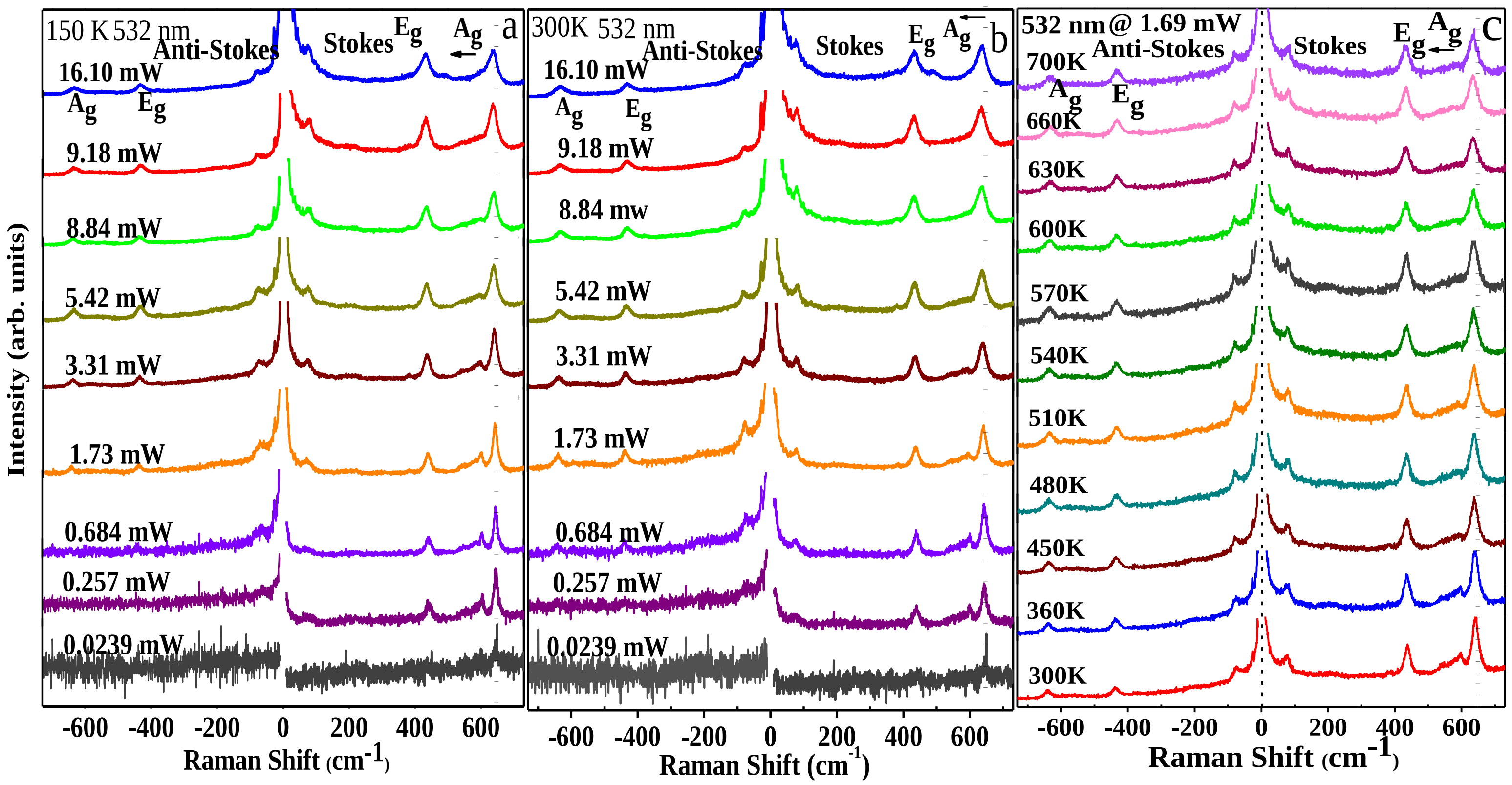

Fig.5

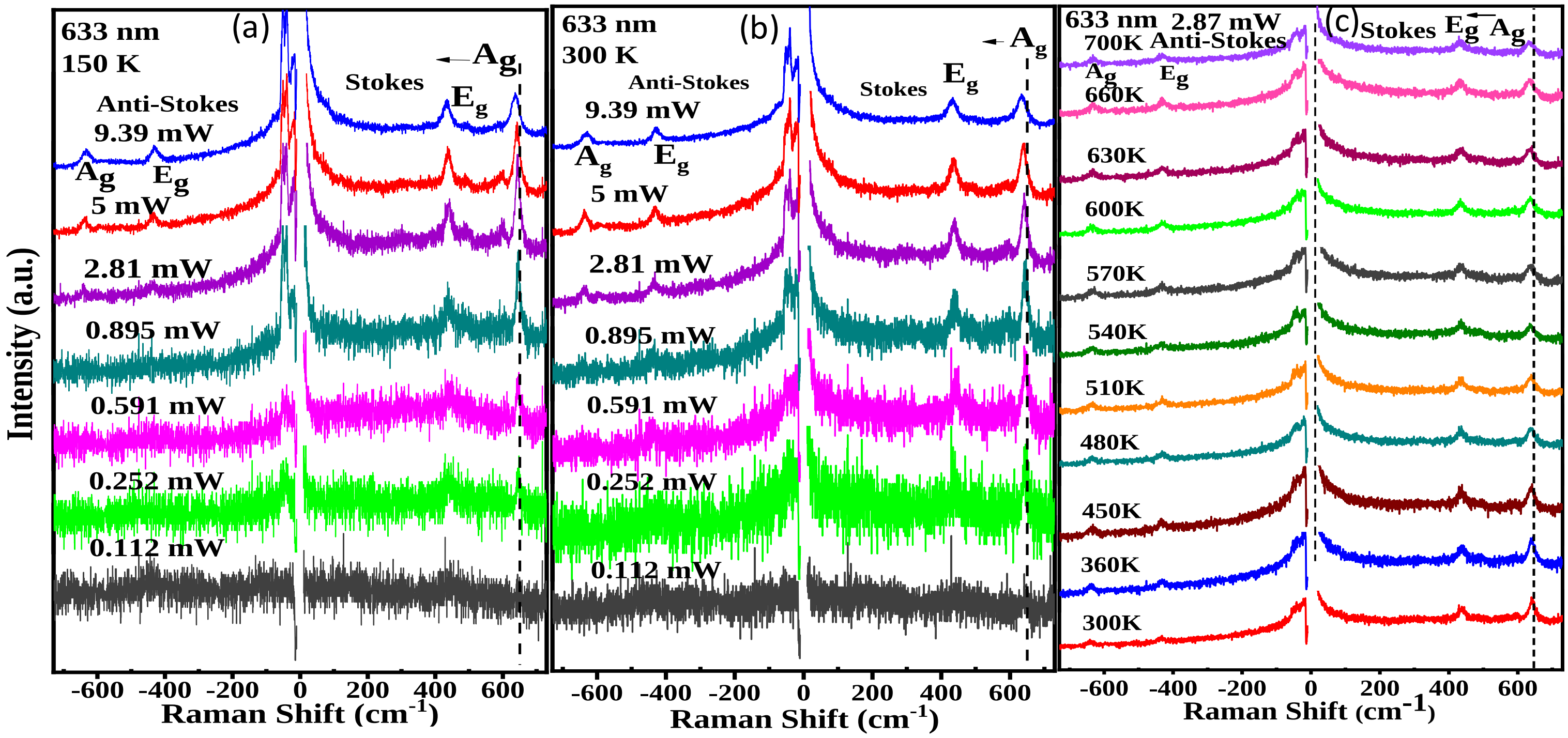

Fig.6

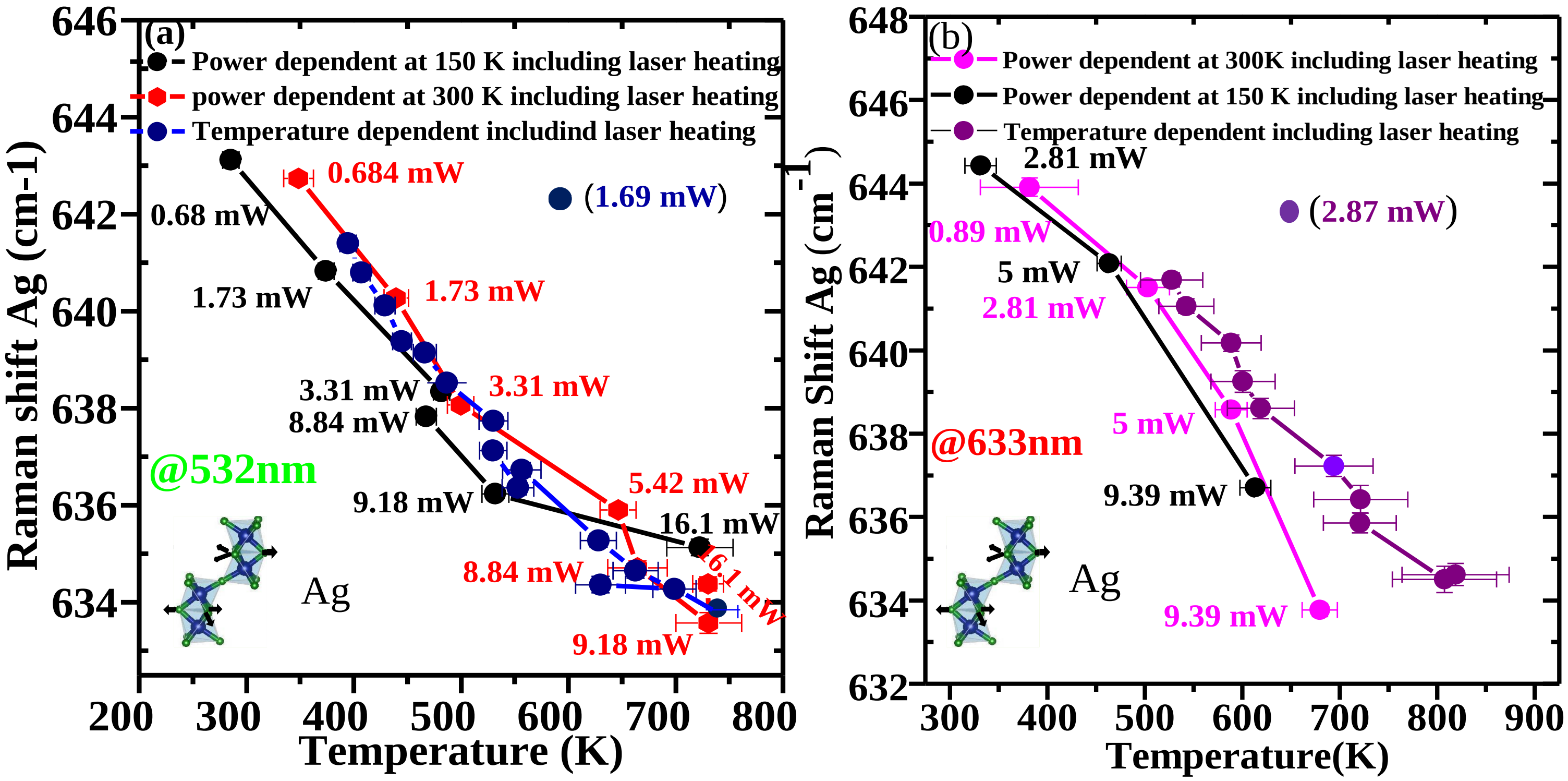

(a)
Power dependent at 150 K including laser heating
power dependent at 300 K including laser heating
Temperature dependent includind laser heating
0.684 mW
(1.69 mW)
0.68 mW
1.73 mW
1.73 mW
3.31 mW
3.31 mW
8.84 mW
@532nm
9.18 mW
5.42 mW
16.1 mW
Ag
8.84 mW
16.1 mW
9.18 mW
Raman shift Ag (cm-1)
Temperature (K)
(b)
Power dependent at 300K including laser heating
Power dependent at 150 K including laser heating
Temperature dependent including laser heating
2.81 mW
(2.87 mW)
0.89 mW
5 mW
2.81 mW
5 mW
@633nm
9.39 mW
Ag
9.39 mW
Raman Shift Ag (cm$^{-1}$)
Temperature(K)

Fig.7

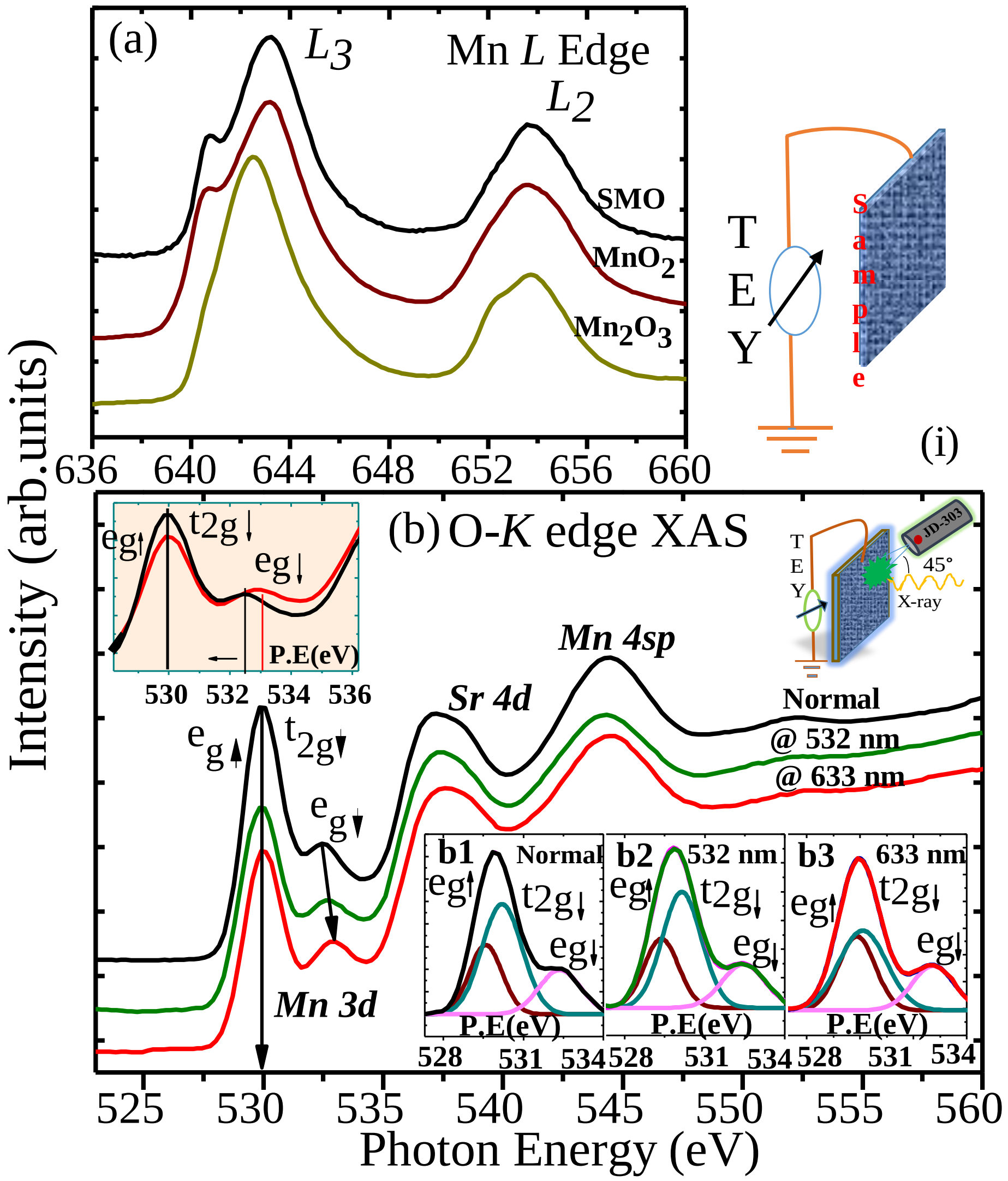

(a)
$L_3$
Mn $L$ Edge
$L_2$
SMO
$MnO_2$
$Mn_2O_3$
636 640 644 648 652 656 660
T
E
Y
Sample
(i)
Intensity (arb.units)
(b) O-$K$ edge XAS
eg↑
t2g↓
eg↓
P.E(eV)
530 532 534 536
JD-303
45°
X-ray
Mn 4sp
Sr 4d
Normal
@ 532 nm
@ 633 nm
$e_g$↑
$t_{2g}$↓
$e_g$↓
Mn 3d
b1 Normal
b2 532 nm
b3 633 nm
528 531 534
Photon Energy (eV)
525 530 535 540 545 550 555 560

Fig.8

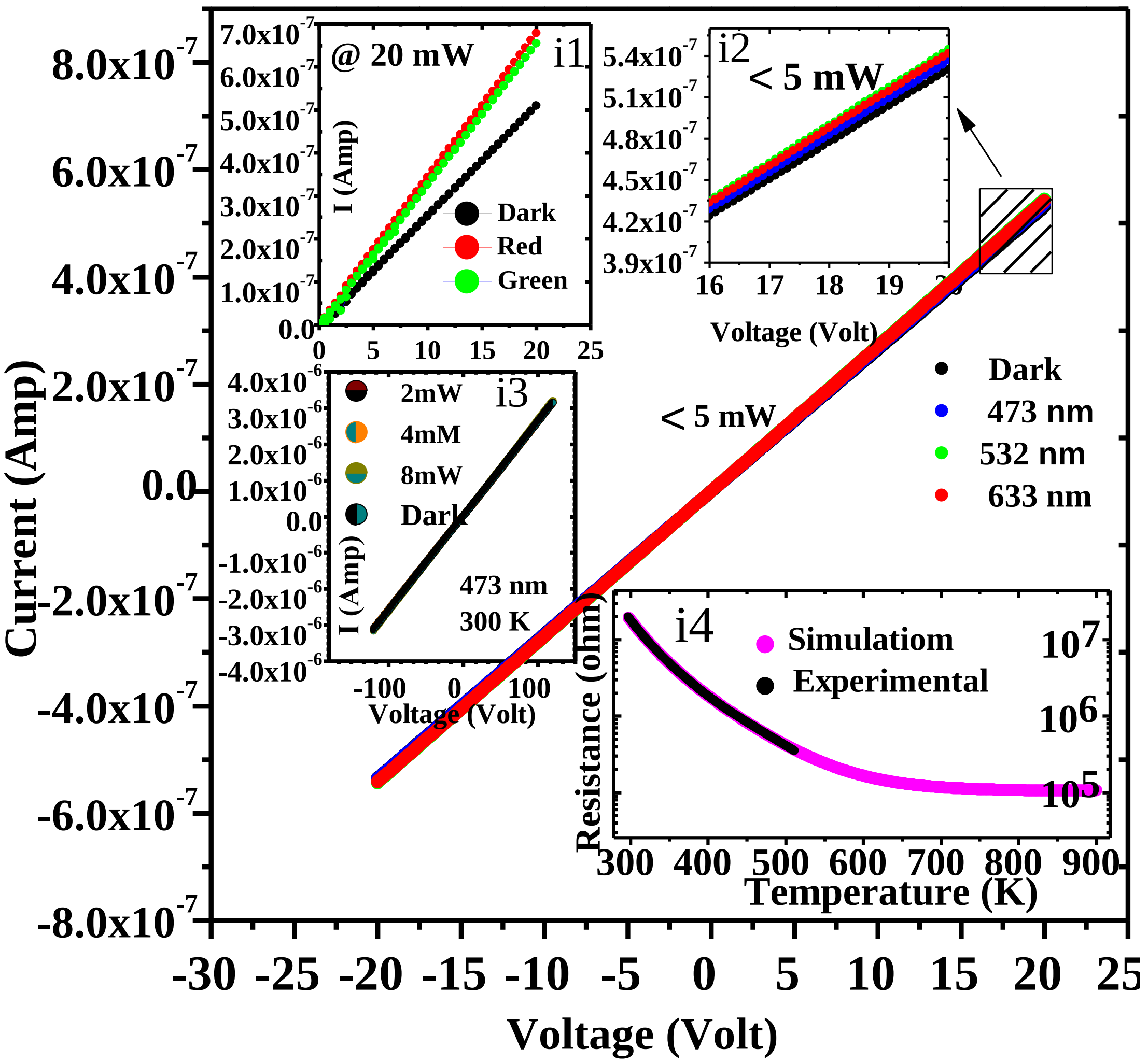

# Photosensitive $SrMnO_3$ (supplementary)

**Arup Kumar Mandal[1], Aprajita Joshi[2] Surajit Saha[2], Binoy Krishna De[1], Sourav Chowdhury[1], VG Sathe[1], U. Deshpande[1], D. Shukla[1], Amandeep Kaur [3], D. M. Phase[1] and R.J. Choudhary[1]***

***[1]UGC-DAE Consortium for Scientific Research, Indore-452001, Madhya Pradesh, India***

***[2]Department of Physics, Indian Institute of Science Education and Research, Bhopal 462066, India***

***[3]Department of Physics, Indian Institute of Technology Bombay, Powai, Mumbai 400076, India***

*Corresponding author: ram@csr.res.in

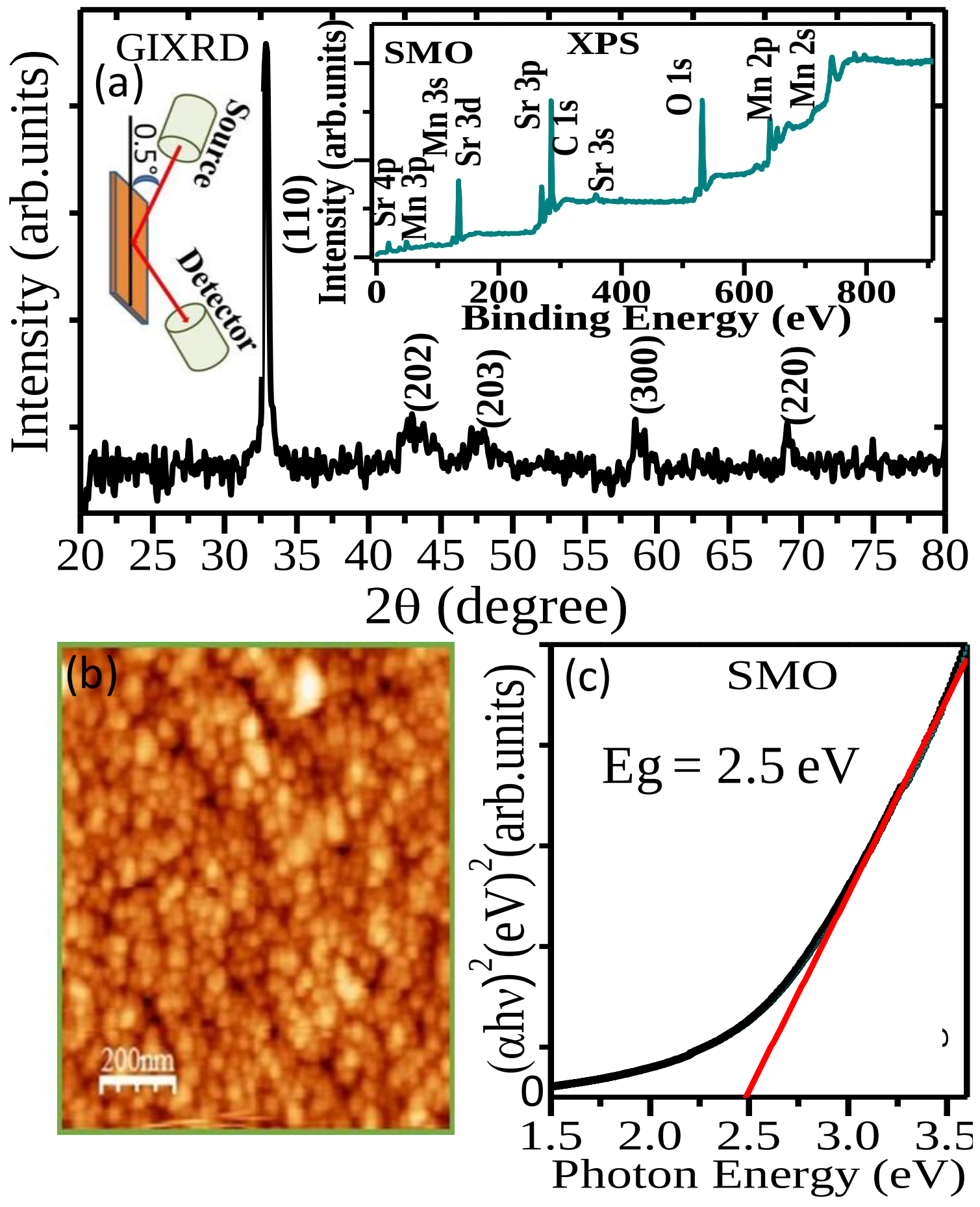

**Fig. S1 (a) GIXRD pattern of 4H-$SrMnO_3$ on quartz, inset figure represents XPS survey scan of the film. (b) AFM image of the grown film, (c) UV-Vis spectra with Tauc fitting.**

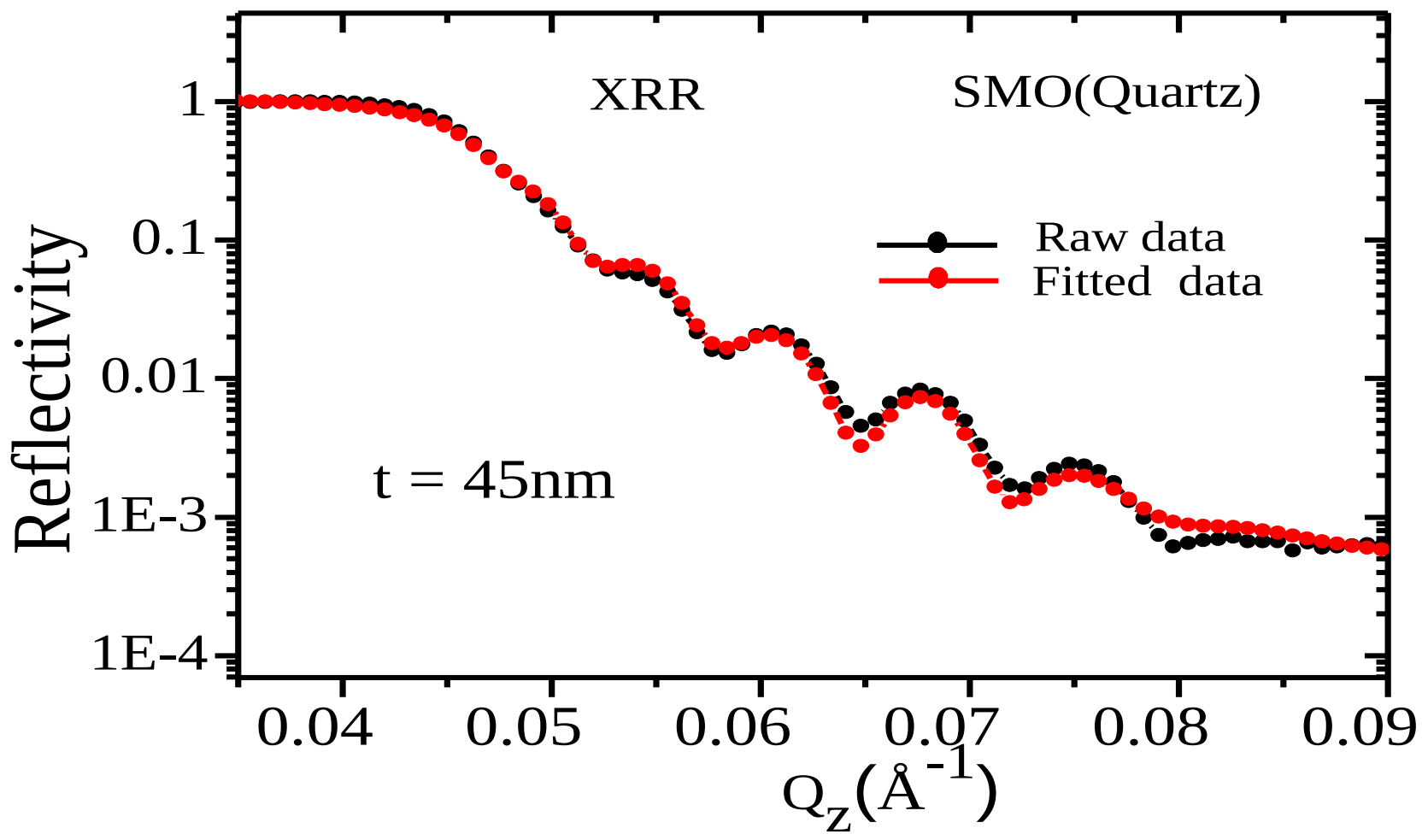

**Fig. S2 X-ray reflectivity of SMO film on quartz substrate.**

Fig S1 (a) represents GIXRD pattern of SMO film on quartz. The pattern reveals Bragg reflections corresponding to (110), (202), (203), (300) and (220) reflections of 4H-$SrMnO_3$ (P63/mmc space group). Calculated lattice parameters of the single phase grown film is a = b = 5.45 Å and c = 9.085Å**.** Fig S1 (a) (inset) represents XPS survey scan of the grown film, revealing that the grown film does not contain any impurity element. For a better understanding of oxygen stoichiometry and chemical charge state of manganese, we performed XAS, and all the details can be found in the main article. Fig S1 (b) shows the AFM image of the grown film recorded in the contact mode, which reveals that the grown film is granular in nature. The measured value of RMS roughness is around 1.2 nm. Fig. S1 (c) is the fitted (Tauc) plot of UV-Vis spectra of the grown film, and the calculated optical band gap of the grown film is 2.5 eV. From XRR, we calculate the film thickness, and it is estimated to be 45 nm (Fig. S2).

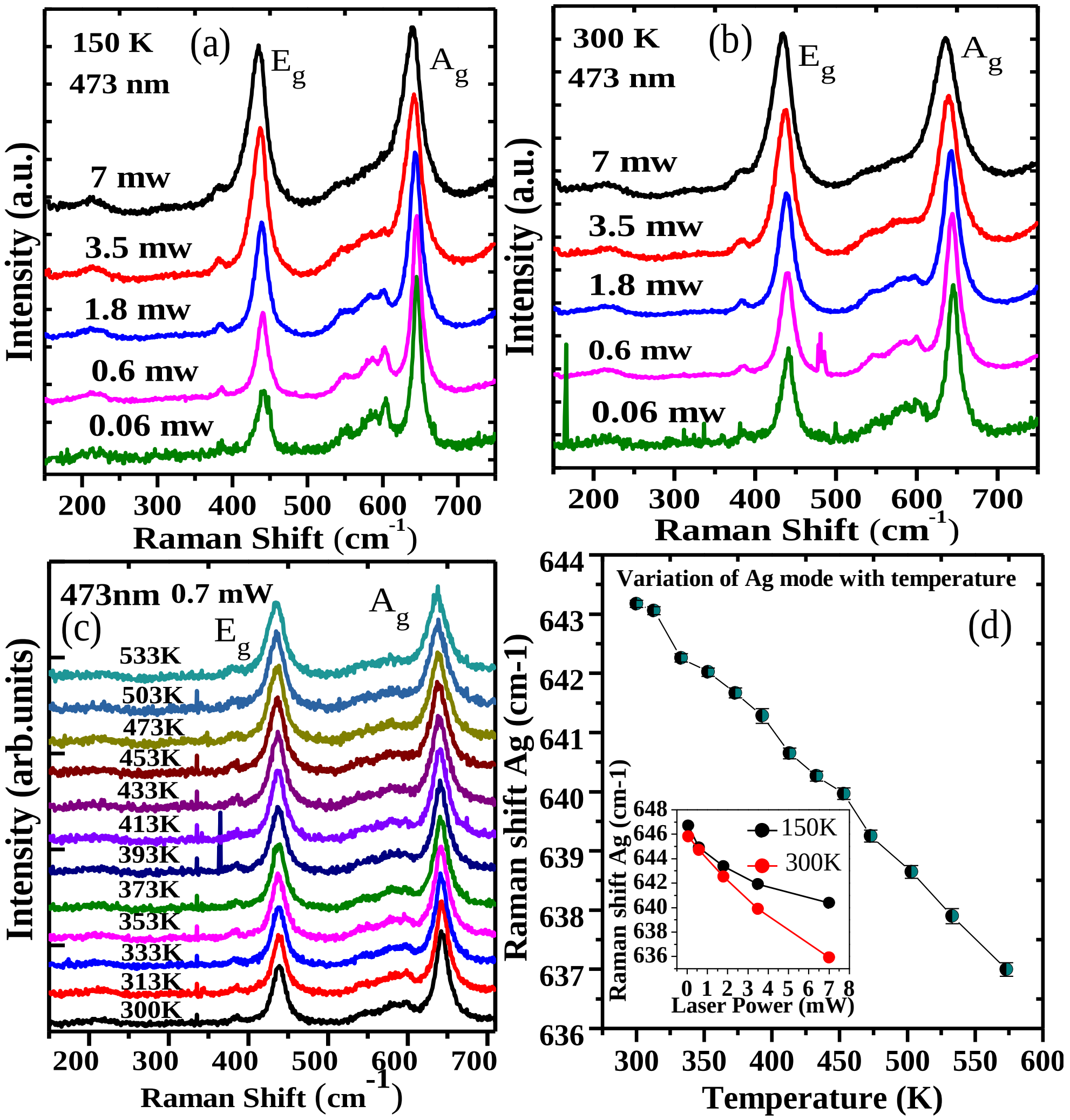

**Fig. S3 (a) Power dependent Raman spectra (stokes line) of SMO film at 150 K with 473 nm laser source. (b) Power dependent Raman spectra of SMO film at 300 K with 473 nm laser source. (c) Temperature dependent Raman spectra with 0.7 mW laser source (473 nm). (d) Variation of Ag mode with temperature; variation of Ag mode with laser power at 150 K and 300 K (Inset)**

Fig. S3 reveals that the most intense Ag mode is shifted towards low frequency with increasing laser power of 473 nm laser source at fixed temperature or with increasing temperature at fixed laser power. Mode softening values are highlighted in the main manuscript.

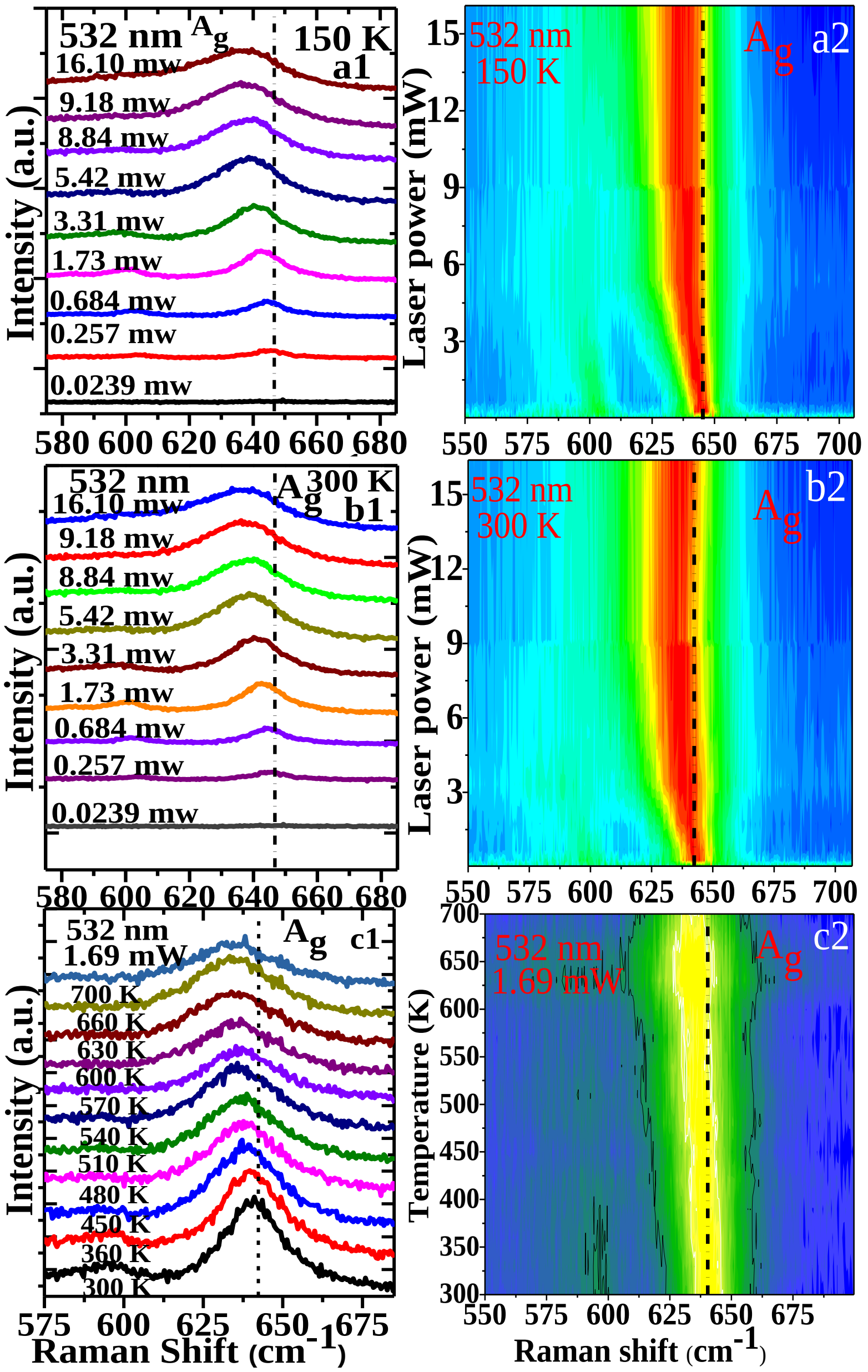

**Fig. S4. (a1), (a2) Power dependent variation of Ag mode at 150 K with 532 nm laser source. (b1), (b2) Power dependent variation of Ag mode at 300 K with 532 nm laser source. (c1), (c2) Temperature dependent variation of Ag mode with 1.69 mW laser source (532 nm).**

From Fig. S4 we conclude that the most intense Ag mode is shifted towards low frequency with 532 nm laser source while increasing laser power at fixed temperature or with increasing temperature at fixed laser power. The value of the mode softening is highlighted in the main manuscript. Full Raman spectra (with stokes and anti-stokes line) are shown in the main manuscript.

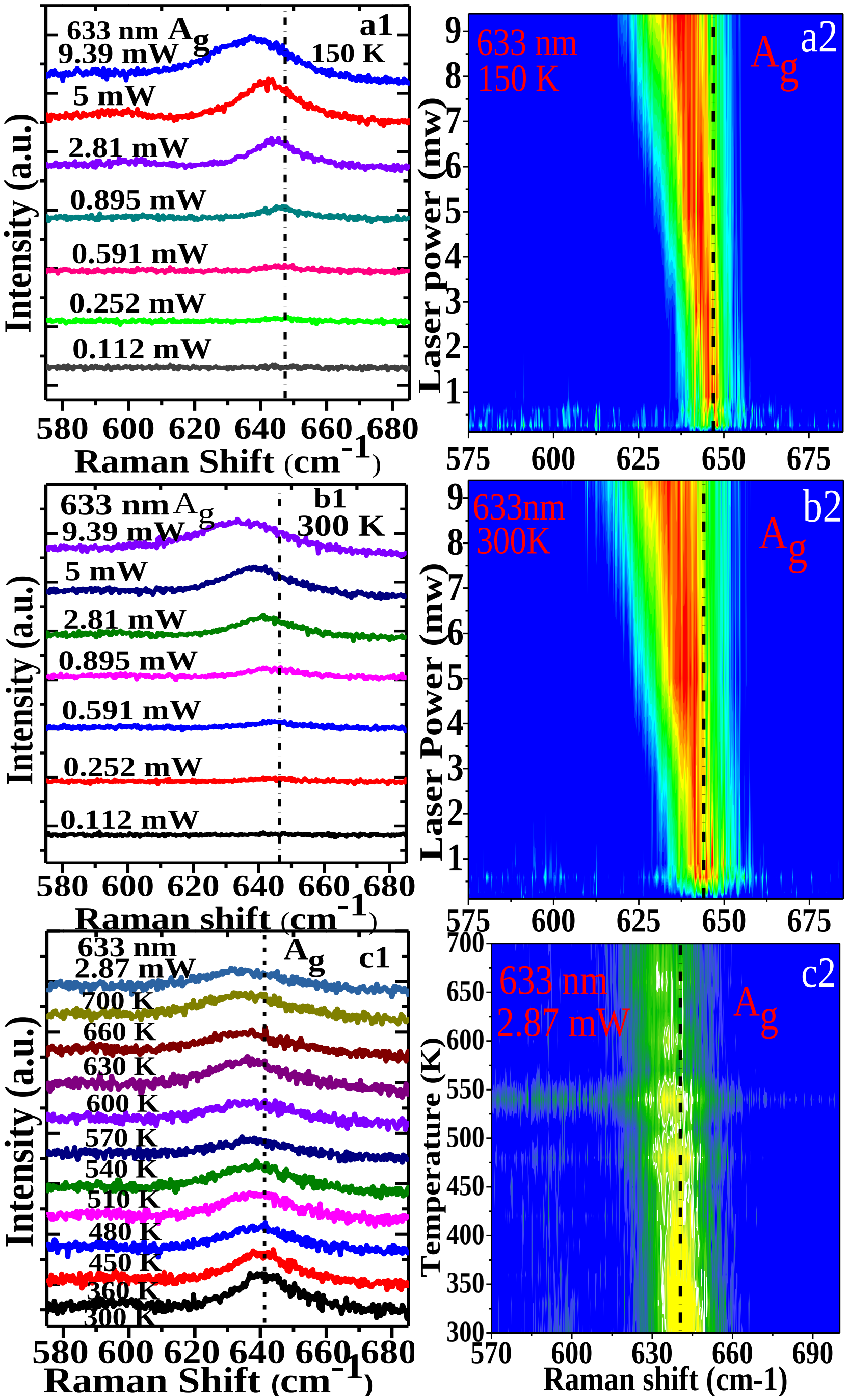

**Fig. S5. (a1), (a2) Power dependent variation of Ag mode at 150 K with 532 nm laser source. (b1), (b2) Power dependent variation of Ag mode at 300 K with 633 nm laser source. (c1), (c2) Temperature dependent variation of Ag mode with 2.87 mW laser source (633 nm).**

Fig. S4 shows that the most intense Ag mode is shifted towards low frequency with 633 nm laser source while increasing laser power at fixed temperature or with increasing temperature at fixed laser power. The value of the mode softening is highlighted in the main manuscript. Full Raman spectra (with stokes and anti-stokes lines) are presented in the main manuscript.